\documentclass[12pt]{article}
\usepackage[margin=2cm]{geometry}
\usepackage{comment}
\usepackage{amsmath,amssymb,extarrows,mathtools,graphicx,subfigure,setspace}
\usepackage{cite}
\usepackage{slashed}
\usepackage{color}
\makeatother

\newcommand{\be}{\begin{equation}}
\newcommand{\bea}{\begin{eqnarray}}
\newcommand{\eea}{\end{eqnarray}}
\newcommand{\ba}{\begin{array}}
\newcommand{\ea}{\end{array}}
\newcommand{\ee}{\end{equation}}
\newcommand{\bes}{\begin{equation*}}
\newcommand{\beas}{\begin{eqnarray*}}
\newcommand{\eeas}{\end{eqnarray*}}
\newcommand{\bas}{\begin{array*}}
\newcommand{\eas}{\end{array*}}
\newcommand{\ees}{\end{equation*}}

\numberwithin{equation}{section}
\begin{document}
\onehalfspacing
\vfill
\begin{titlepage}
\vspace{10mm}
\begin{flushright}
\end{flushright}
  %
\vspace*{20mm}
\begin{center}
{\Large {\bf  Born-Infeld Gravity Revisited}\\
}

\vspace*{15mm}
\vspace*{1mm}
{M. R. Setare\footnote{e-mail:rezakord@ipm.ir} and M. Sahraee \footnote{e-mail:m.sahraee@uok.ac.ir}}

 \vspace*{1cm}

{\it  Department of Science, Campus of Bijar, University of  Kurdistan  \\
Bijar, IRAN. }

 \vspace*{0.5cm}

\vspace*{2cm}
\end{center}

\begin{abstract}
In this paper we investigate the behavior of linearized gravitational excitation in the Born-Infeld Gravity in $AdS_3$ space.
We obtain the linearized equation of motion and show that this higher order gravity propagate two gravitons, massless and massive, on the $AdS_3$ background. In contrast to the $R^2$ models, such as TMG or NMG,  Born-Infeld Gravity does not have a critical point for any regular choice of parameters. So the logarithmic solution is not a solution of this model, due to this one can not find a logarithmic conformal field theory as a dual model for Born-Infeld Gravity.

\hspace{0.35cm}


\end{abstract}

\end{titlepage}

\section{Introduction}
It is well known that Einstein gravity suffers from the problem that
the theory is nonrenormalizable in four and higher dimensions.
Adding higher derivative terms such as Ricci and scalar curvature
squared terms makes the theory renormalizable at the cost of the
loss of unitarity \cite{a}. A few years ego \cite{b} a new theory of massive gravity (NMG) in three dimensions
has been proposed. This theory is
 equivalent to the three-dimensional
Fierz-Pauli action for a massive spin-2 field at the linearized level. Moreover NMG in contrast with the Topologically
Massive Gravity (TMG) \cite{c} is parity
invariant.  With the only Einstein-
Hilbert term in the action there are no propagating degrees of freedom, but by adding  the higher curvature terms in the action the situation becomes different. Usually the theories including the
terms given by the square of the curvatures have the massive spin 2
mode and the massive scalar mode in addition to the massless
graviton. Also the theory has ghosts due to negative energy
excitations of the massive tensor. The unitarity of NMG
 was discussed in \cite{d1,d2} (see also \cite{dan11,dan12}) and this model is generalized to higher dimensions. There are many works on new theory of massive gravity, see for example \cite{e1,e2,e3,e4,e5,e6}. Recently,
quadratic-curvature actions with cosmological constant have been
introduced in four \cite{f} and $D$ \cite{g} dimensions. It was
found that there exist a choice of parameters for which these
theories possess one AdS background on which neither massive fields,
nor massless scalars propagate. By this special choice of the
parameters, which is called as a critical point, there appears a
mode which behaves as a logarithmic function of the distance. The massive graviton modes obey Brown-Henneaux boundary conditions, at the critical point in parameter space, the massive gravitons become massless and are replaced by new modes, so-called the logarithmic modes. Logarithmic
modes in the framework of the higher-dimensional critical gravity models
were recently found in \cite{gl}. After that Bergshoeff et. al \cite{ber} have obtained the logarithmic solutions of the linearized
$4$-dimensional critical gravity, they have shown
that logarithmic modes are of two types, spin 2 and Proca log modes. The
number of independent spin 2 log modes is given by the number of polarization
states of a massless spin 2 field, while the number of independent
Proca log modes is given by the number of polarization states of a massive spin 1 field.\\
  There is this possibility to extend NMG to higher curvature theories. One of these extension
 of NMG has been done by Sinha \cite{2} where he has added the $R^3$ terms to the action. The other modification is the extension to the Born-Infeld type action \cite{h}. The (warped) AdS black hole solutions for these extensions of NMG have been obtained in \cite{1}.\\ In this paper we would like to investigate the behavior of the linearized gravitational excitations of
Born-Infeld gravity in the background of
$AdS_3$ spacetime with Brown-Henneaux boundary conditions. The similar study for TMG and NMG previously have been done respectively in \cite{3,4}.
More than this may be one ask about log modes in critical case of Born-Infeld gravity. Because as we mentioned, this
is a subject of current interest, since log modes are present in the so-called critical theories of gravity which makes these theories possibly dual to log CFTs. But in contrast to TMG and NMG models, here we show that in the framework of  Born-Infeld gravity and in any regular choice of parameters, there is not any critical point. So logarithmic modes do not arises and we can not find a logarithmic conformal filed theory as dual model for  Born-Infeld gravity.\\
 Following \cite{h,5}, a type of extending NMG is based on a Born-Infeld type action. It is interesting that this action at the first order of the curvature gives the cosmological Einstein-Hilbert action, it gives the NMG action at the second order of the curvature and have been derived that at the third and fourth orders, it gives the action presented in \cite{2}. The Born-Infeld type extension of (non)-critical gravity on AdS space has been considered in \cite{yi}, in this paper Yi has shown that this extension is consistent with holographic C-theorem. In reference \cite{tek} the unitarity analysis of general Born-Infeld gravity theories around
their constant curvature vacua have been done (see also \cite{tek2}).

 The authors of \cite{js} have shown that the Born-Infled extension of NMG
 emerges naturally as a counterterm in $AdS_4$.  Here we consider generally the Born-Infeld action \cite{5}. It was shown in \cite{sin} that in the context of critical gravities arising from  Born-Infeld type actions, the trace of the fluctuation (scalar) $h$ becomes dynamical.
In fact unlike $R^2$ theories where the equations of motion set $h$
to zero \cite{f,g,3,4} in these theories it is not necessary to set h=0.\\
This paper is organized as follows. In section 2, we review briefly Born-Infeld Gravity. Then in section 3, we obtain the equation of motion for the linearized excitation $h_{\mu\nu}$ around the $AdS_{3}$ background solution. After that in section 4, we obtain the linearized solutions  of the model. In section 5, we calculate the energy of  the linearized
gravitons in $AdS_3$ background. Our results are summarized
 in section 6.

\section {The Born-Infeld Gravity}
In this section we review the Born-Infeld action and its equation of motion. The action for the Born-Infeld Gravity can be written as \cite{h,5} \begin{equation}
I_{BI-NMG}=-\frac{4m^{2}}{\kappa^2}\int d^{3}x[\sqrt{-det(g+\frac{\sigma}{m^{2}}G)}-(1-\frac{\lambda}{2})\sqrt{-det g}],
\end{equation}
where $m$ is mass parameter, $\lambda$ is also the cosmological constant term and $\sigma$ take $1$  or $-1$. Here $G_{\mu\nu}$ is given by  \begin{equation}G_{\mu\nu}\equiv R_{\mu\nu}-\frac{1}{2}g_{\mu\nu}R \end{equation}In \cite{5} has been shown that the above action reduces to the following action \begin{equation}I_{BI-NMG}=-\frac{4m^{2}}{\kappa^2}\int d^{3}x\,\sqrt{-det g
} F(R,K,S),\end{equation}
where \begin{equation}F(R,K,S)\equiv\sqrt{1-\frac{\sigma}{2 m^{2}}(R+\frac{\sigma}{m^{2}}K-\frac{1}{12 m^{4}}S)}-(1-\frac{\lambda}{2}),\end{equation}  and\begin{equation}K=R_{\mu\nu}R^{\mu\nu}-\frac{1}{2}R^2,\,\,\,\,\,\,\,\,\,\,\,\,\,\,\,\,\,\,\,\,\,\,\,\,\,\,\,\,\,\,\,\,\,\,\,\,\,\,\,\,\,\,\,\,\,\,\end{equation}
\begin{equation}S\equiv8R^{\mu\nu}R_{\mu\alpha} R_{\nu}^{\alpha}-6RR_{\mu\nu}R^{\mu\nu}+ R^{3}.\,\,\,\,\end{equation}
 We assume that $\kappa^{2}$, $m^{2}$ are always positive.
 With choosing $\sigma=-1$, the equation of motion of this action is given\cite{5} \begin{equation}-\frac{\kappa^{2}}{8m^{2}}T_{\mu\nu}=-\frac{1}{2}Fg_{\mu\nu}+
 (g_{\mu\nu}\Box-\nabla_{\mu}\nabla_{\nu})F_{R}+F_{R}R_{\mu\nu}\,\,\,\,\,\,\,\,\,\,\,\,\,\,\,\,\,\,\,\,\,\,\,\,\,\,\,\,\,\,\,\,\,\,\,\,\,\,\,\,\,\,\,\,\,\,\,
 \,\,\,\,\,\,\,\,\,\,\,\,\,\,\,\,\,\,\,\,\,\,\,\,\,\,\,\,\,\,\,\,\,\,\,\,\,\,\,\,\,
  \,\,\,\,\,\,\,\,\,\,\,\,\,\,\,\,\,\,\,\,\,\,\,\,\,\,\,\nonumber\end{equation} \begin{equation}+\frac{1}{m^{2}}F_{R}[RR_{\mu\nu}-2R_{\lambda\nu\alpha\mu}R^{\lambda\alpha}-\Box(R_{\mu\nu}-\frac{1}{2}g_{\mu\nu}R)]
\,\,\,\,\,\,\,\,\,\,\,\,\,\,\,\,\,\,\,\,\,\,\,\,\,\,\,\,\,\,\,\,\,,\,\,\,\,\,\,\,\,\,\,\,\,\,\,\,\,\,\,\,\,\,\,\,\,\,\nonumber\end{equation}\begin{equation}+\frac{1}{m^{2}}[2\nabla_{\alpha}F_{R}\nabla_{\mu}R_{\nu}^{\alpha}-
2\nabla_{\alpha}F_{R}\nabla^{\alpha}
(R_{\mu\nu}-\frac{1}{2}g_{\mu\nu}R)-\nabla_{\nu}F_{R}\nabla_{\mu}R]\,\,\,\,\,\,\,\,\,\,\,\,\,\,\,\,\,\,\,\,\,\,\,\,\,\,\,\,\nonumber\end{equation}\begin{equation}+\frac{1}{m^{2}}[(2R_{\nu}^{\alpha}\nabla_{\alpha}
\nabla_{\mu}-g_{\mu\nu}R^{\alpha\beta}\nabla_{\beta}\nabla_{\alpha}-R_{\mu\nu}\Box)+R(g_{\mu\nu}\Box-\nabla_{\mu}\nabla_{\nu})]F_{R}\,\,\,\,\,\,\,\,\,\,
\,\,\,\,\,\nonumber\end{equation}
\begin{equation}-\frac{1}{m^{4}}F_{R}(2R_{\mu}^{\rho}R_{\rho\alpha}R_{\nu}^{\alpha}+[g_{\mu\nu}\nabla_{\alpha}\nabla_{\beta}(R^{\beta\rho}R_{\rho}^{\alpha})+\Box
(R_{\nu}^{\rho}R_{\mu\rho})-2\nabla_{\alpha}\nabla_{\mu}(R_{\nu}^{\rho}R_{\rho}^{\alpha}])\nonumber\end{equation}
\begin{equation}-\frac{1}{2m^{2}}F_{R}([2\nabla_{\alpha}\nabla_{\mu}(RR_{\nu}^{\alpha})-g_{\mu\nu}\nabla_{\alpha}\nabla_{\beta}(RR^{\alpha\beta})-
\Box(RR_{\mu\nu})]-2RR_{\nu}^{\rho}R_{\mu\rho})\,\,\,\,\nonumber\end{equation}\begin{equation}+\frac{1}{2m^{4}}F_{R}[(g_{\mu\nu}\Box-\nabla_{\nu}\nabla_{\mu})+
R_{\mu\nu}](R_{\mu\nu}^{2}-\frac{1}{2}R^{2})\,\,\,\,\,\,\,\,\,\,\,\,\,\,\,\,\,\,\,\,\,\,\,\,\,\,\,\,\,\,\,\,\,\,\,\,\,\,\,\,\,\,\,\,\,\,\,\,\,\,\,\,
\,\,\,\,\,\,\,\,\,\,\,\,\,\,\,\,\nonumber\end{equation}\begin{equation}-\frac{2}{m^{4}}\nabla_{\alpha}F_{R}[\nabla^{\alpha}(
R_{\nu}^{\rho}R_{\mu\rho})+g_{\mu\nu}\nabla_{\beta}(R^{\beta\rho}R_{\rho}^{\alpha})-\nabla_{\mu}(R_{\nu}^{\rho}R_{\rho}^{\alpha})]
\,\,\,\,\,\,\,\,\,\,\,\,\,\,\,\,\,\,\,\,\,\,\,\,\,\,\,\,\,\,\,\,\,\,\,\,\,\,\,\,\,\,\,\,\,\nonumber\end{equation}\begin{equation}
+\frac{2}{m^{4}}\nabla_{\alpha}(R_{\nu}^{\rho}R_{\rho}^{\alpha})\nabla_{\mu}F_{R}+\frac{1}{m^{4}}\nabla_{\alpha}F_{R}[\nabla_{\alpha}(RR_{\mu\nu})
+g_{\mu\nu}\nabla_{\beta}(RR^{\alpha\beta})-\nabla_{\mu}(RR_{\nu}^{\alpha})]\nonumber\end{equation}\begin{equation}-\frac{1}{m^{4}}\nabla_{\alpha}
(RR_{\nu}^{\alpha})\nabla_{\alpha}F_{R}-\frac{1}{2m^{4}}(\nabla_{\mu}F_{R}\nabla_{\nu}+\nabla_{\nu}F_{R}\nabla_{\mu}-2g_{\mu\nu}\nabla_{\rho}F_{R}
\nabla^{\rho})(R_{\alpha\beta}^{2}-\frac{1}{2}R^{2})\nonumber\end{equation}
\begin{equation}-\frac{1}{m^{4}}(g_{\mu\nu}R^{\beta\rho}R_{\rho}^{\alpha}\nabla_{\alpha\alpha}\nabla_{\beta}+
R_{\nu}^{\rho}R_{\mu\rho}\Box-2R_{\nu}^{\rho}R_{\rho}^{\alpha}\nabla_{\alpha}\nabla_{\mu})F_{R}\,\,\,\,\,\,\,\,\,\,\,\,\,\,
\,\,\,\,\,\,\,\,\,\,\,\,\,\,\,\,\,\,\,\,\,\,\,\,\,\,\,\,\,\,\,\,\,\,\,\,\,\,\,\,\,\,\nonumber\end{equation}\begin{equation}-\frac{1}{2m^{4}}(2RR_{\nu}^{\alpha}\nabla
_{\alpha}\nabla_{\mu}-g_{\mu\nu}RR^{\alpha\beta}\nabla_{\alpha}\nabla_{\beta}-RR_{\mu\nu}\Box)F_{R}\,\,\,\,\,\,\,\,\,\,\,\,\,\,\,\,\,\,\,\,\,\,\,\,
\,\,\,\,\,\,\,\,\,\,\,\,\,\,\,\,\,\,\,\,\,\,\,\,\,\,\,\,\,\,\,\,\,\,\,\,\,\,\,\,\,\,\nonumber\end{equation}
\begin{equation}+\frac{1}{2m^{4}}(R_{\alpha\beta}^{2}-\frac{1}{2}R^{2})(g_{\mu\nu}\Box-\nabla_{\nu}\nabla_{\mu})F_{R},
\,\,\,\,\,\,\,\,\,\,\,\,\,\,\,\,\,\,\,\,\,\,\,\,
\,\,\,\,\,\,\,\,\,\,\,\,\,\,\,\,\,\,\,\,\,\,\,\,
\,\,\,\,\,\,\,\,\,\,\,\,\,\,\,\,\,\,\,\,\,\,\,\,
\,\,\,\,\,\,\,\,\,\,\,\,\,\,\,\,\,\,\,\,\,\,\,\,\,\,\,\,
\end{equation}where \begin{equation}F_{R}=\frac{\partial F}{\partial R}=\frac{1}{4m^{2}[F+(1-\frac{\lambda}{2})]}.\end{equation}
with a non-zero cosmological constant $\Lambda=-\frac{1}{\ell^{2}}$. The Born-Infeld Gravity admit an $AdS_{3}$ solution \begin{equation}dS^{2}=\bar{g}_{\mu\nu}dx^{\mu}dx^{\nu}=\ell^{2}(-\cosh^{2}\rho d\tau^{2}+\sinh^{2}\rho d\phi^{2}+ d\rho^{2}),\end{equation}
the Riemann tensor, Ricci tensor and Ricci scalar of the above $AdS_3$ metric are\begin{equation}\bar{R}_{\mu\alpha\upsilon\beta}=\Lambda(\bar{g}_{\mu\nu}\bar{g}_{\alpha\beta}-\bar{g}_{\mu\beta}\bar{g}_{\alpha\nu}),\,\,\,\,\,\,\,
\bar{R}_{\mu\nu}=2\Lambda\bar{g}_{\mu\nu},\,\,\,\,\,\,\,\,  \bar{R}=6\Lambda\end{equation}with substituting these in Eq.(2.4), F become
\begin{equation}F(R,K,S)=\sqrt{(1+\frac{\Lambda}{m^{2}})^{3}}-(1-\frac{\lambda}{2}),\end{equation}putting these in the equation of motion (2.7), one obtains
\begin{equation}0=(-\frac{1}{2}F+\frac{\Lambda}{2m^{2}[F+(1-\frac{\lambda}{2})]}+\frac{\Lambda^{2}}{m^{4}[F+(1-\frac{\lambda}{2})]}+\frac{\Lambda^{3}}{2m^{6}[F+(1-\frac{\lambda}{2})]
})g_{\mu\nu}.\end{equation}By solving (2.12), the parameter $\lambda$ should be related to the cosmological constant $\Lambda$ and the mass parameter as follows \begin{equation}\Lambda=-m^{2}\lambda(1-\frac{\lambda}{4}),\,\,\,\,\,\,\,\,\,\,\,\,\,\,\lambda<2\end{equation}
Note that the inequality $\lambda<2$, can not be saturated, since $\lambda=2$ would imply $\Lambda=-m^{2}$ and thereby lead to vanishing $F$, however, Eq.(2.13) was derived assuming that $F$ is non-vanishing (and real).

\section{Gravitons in $AdS_{3}$ }
By expanding $g_{\mu\nu}=\bar{g}_{\mu\nu}+h_{\mu\nu}$ around the $AdS_{3}$ background solution $(2.9)$, one can obtain the equation of motion for the linearized excitation $h_{\mu\nu}$ as following \begin{equation}\bar{\Box}\bar{\Box} h_{\mu\nu}+ A\bar{\Box} h_{\mu\nu
}+ B h_{\mu\nu}=0,\end{equation}
 where  the traceless and divergenceless conditions on $h_{\mu\nu}$  have been imposed, i.e., $\bar{\nabla}_{\mu}h^{\mu\nu}=h=0$ and $h_{\mu}^{\mu}=h=0$, respectively. A and B are given by
 \begin{equation}A=\frac{1}{m^{2}+\Lambda}(-m^{4}-6\Lambda m^{2}-5\Lambda^{2}),\end{equation}
 \begin{equation}B=\frac{1}{m^{2}+\Lambda}(2\Lambda m^{4}+8\Lambda^{2}m^{2}+6\Lambda^{3}).\end{equation}The above equation could be factorized as \begin{equation}(\bar{\Box}-2\Lambda)(\bar{\Box}-2\Lambda-M^{2})h_{\mu\nu}=0,\end{equation}where\begin{equation}M^{2}=m^{2}+\Lambda.  \end{equation} This equation has solutions that correspond to a massless mode $h_{\mu\nu}^{m}$ and a massive mode $h_{\mu\nu}^{M}$ that satisfy the following equations of motion:
\begin{equation}(\bar{\Box}-2\Lambda)h_{\mu\nu}^{m}=0,\,\,\,\,\,\,\,\,\,\,\,\,\,\,\,   (\bar{\Box}-2\Lambda -M^{2})h_{\mu\nu}^{M}=0.\end{equation}

 \section{Linearized solutions }
 We know that the isometry group of $AdS_{3}$ space is $SL(2,R)_{L}\times SL(2,R)_{R}$, so one can write the Laplacian acting on tensor $h_{\mu\nu}$ in terms of the sum of Casimir operators of $SL(2,R)_{L}$ and $SL(2,R)_{L}$, \cite{3}, (see also \cite{dan1,dan2})
  \begin{equation}\bar{\nabla}^{2}h_{\mu\nu}=-[\frac{2}{\ell^{2}}(L^{2}+\bar{L}^{2})+\frac{6}{\ell^{2}}]h_{\mu\nu}.\end{equation} So the equation of motion $(3.4)$ can be rewritten as  \begin{equation}[-\frac{2}{\ell^{2}}(L^{2}+\bar{L}^{2})-\frac{6}{\ell^{2}}+\frac{2}{\ell^{2}}][-\frac{2}{\ell^{2}}(L^{2}+\bar{L}^{2})-\frac{6}{\ell^{2}}+\frac{2}{\ell^{2}}-M^{2}]h_{\mu\nu}=0,\end{equation} By considering the highest weight states, where $L^{2}|\psi_{\mu\nu}\rangle=-h(h-1)|\psi_{\mu\nu}\rangle$, from Eq.$(4.2)$ we can write  \begin{equation}[h(h-1)+\bar{h}(\bar{h}-1)-2][h(h-1)+\bar{h}(\bar{h}-1)-2-\frac{1}{2}M^{2}\ell^{2}]h_{\mu\nu}=0,\,\, h-\bar{h}=\pm2\end{equation}There are two branches of solutions for the above equation. For the massless mode we have  \begin{equation}h(h-1)+\bar{h}(\bar{h}-1)-2=0,\end{equation}for $h=2+\bar{h}$, we obtain \begin{equation}\bar{h}=\frac{-1\pm1}{2},\,\,\,\,\,\,\,\,\,   h=\frac{3\pm1}{2}\end{equation}but for $h=-2+\bar{h}$ , we have \begin{equation}\bar{h}=\frac{3\pm1}{2},\,\,\,\,\,\,\,\,\,    h=\frac{-1\pm1}{2}.\end{equation}Now for second
  branch i.e, the second massive mode, \begin{equation}h(h-1)+\bar{h}(\bar{h}-1)-2-\frac{1}{2}M^{2}\ell^{2}=0,\end{equation} for $h=2+\bar{h}$, we have \begin{equation}\bar{h}=\frac{-1\pm\sqrt{1+M^{2}\ell^{2}}}{2},\,\,\,\,  \,\,\,\,\,\  h=\frac{3\pm\sqrt{1+M^{2}\ell^{2}}}{2}\end{equation}and for $h=-2+\bar{h}$, we have  \begin{equation}\bar{h}=\frac{3\pm\sqrt{1+M^{2}\ell^{2}}}{2},\,\,\,\,  \,\,\,\,\,\  h=\frac{-1\pm\sqrt{1+M^{2}\ell^{2}}}{2}.\end{equation}
  As have been mentioned
in \cite{1,3}, the solutions with the
  lower sign will blow up at infinity, thus we consider only the ones with the upper sign, i.e $ (\frac{-1+\sqrt{1+M^{2}\ell^{2}}}{2},\frac{3+\sqrt{1+M^{2}\ell^{2}}}{2})$ and $(\frac{3+\sqrt{1+M^{2}\ell^{2}}}{2},\frac{-1+\sqrt{1+M^{2}\ell^{2}}}{2})$.\\
  In contrast to the NMG, TMG, and GMG, the Born-Infeld gravity does not have a critical point for any regular choice of parameters. The critical point can be obtain when $M^2=0$, this is the tuning required for the presence of logarithmic modes. But $M^2=0$, with Eq.(3.8) implies $m^2=-\Lambda$, which in turn implies through Eq.(2.13) the equality $\lambda=2$. In other words, logarithmic modes are present only at $\lambda=2$. But this is precisely the value that is already forbidden by the analysis in the end of section 2. In the other hand may be one look for a possible limiting solution when $M^{2}\ell^{2}\rightarrow0$ then results  $m^{2}\ell^{2}\rightarrow1$.  Therefore only there is a limiting critical point $m^{2}\ell^{2}\rightarrow1$ in the parameter space $m^{2}$, $\Lambda$; whereas we have $m^{2}>0$ and $\ell^{2}>0$ (since $\ell^{2}=-\frac{1}{\Lambda}$ and $\Lambda$ always is negative in the $AdS_{3}$ background).  Because in this limiting point the above massive modes take the form $(0,2)$ and $(2,0)$ which are correspond to the massless modes and the energies of both highest weight massless and massive modes are zero. When two linearly independent solutions to a differential equation degenerate, a logarithmic
solution appears. This is similar to what have been done in \cite{ali} for AdS wave solutions, where the authors did derive the equations for BI gravity
at the critical point as a limit of the non-critical theory. So as a limiting
case, the pure BI gravity theory is indeed critical, and we obtain a limiting logarithmic solution. But if you
set $\lambda=2$, this will not be valid as we have mentioned, and it can be considered as a limiting solution.
 \section{The energy of linearized gravitons  }
 Now we would like to calculate the energy of the linearized gravitons in $AdS_{3}$ background. The fluctuation $h_{\mu\nu}$ can be decomposed as\cite{3,4} \begin{equation}h_{\mu\nu}=h_{\mu\nu}^{m}+h_{\mu\nu}^{M}\end{equation} Up to total derivatives, the quadratic action of $h_{\mu\nu}$ is given by \begin{equation}I_{2}=\frac{m^{2}+\Lambda}{m^{2}\kappa^2}\int d^{3}x\sqrt{-\bar{g}}(\bar{\Box}\bar{\Box} h_{\mu\nu}+ A\bar{\Box} h_{\mu\nu
}+ B h_{\mu\nu}) h^{\mu\nu}\nonumber\,\,\,\,\,\,\,\,\,\,\,\,\,\,\,\,\,\,\,\,\,\,\,\,\,\,\,\,\,\,\,\end{equation}\begin{equation}=\frac{m^{2}+\Lambda}{m^{2}\kappa^2}\int d^{3}x \sqrt{-\bar{g}}
  (\bar{\Box}h^{\mu\nu}\bar{\Box}h_{\mu\nu} - A \bar{\nabla}_{\lambda}h_{\mu\nu}\bar{\nabla}^{\lambda}h^{\mu\nu}+ B h_{\mu\nu}h^{\mu\nu}).\end{equation}The momentum conjugate to $h_{\mu\nu}$ is\begin{equation}\Pi^{(1)\mu\nu}=\frac{\delta \pounds}{\delta(\bar{\nabla}_{0}h_{\mu\nu})}=-\frac{2(m^{2}+\Lambda)}{m^{2}\kappa^2}\sqrt{-g} [\bar{\nabla}^0\bar{\Box} h^{\mu\nu}+ A \bar{\nabla}^0 h^{\mu\nu}],\end{equation}where $\pounds$ is Lagrangian density.

  Now using the equation of motion we can obtain following expression for momentum conjugate to the massive modes $h_{\mu\nu}^{m}$ and $h_{\mu\nu}^{M}$
  respectively\begin{equation}\Pi_{m}^{(1)\mu\nu}=-\frac{2(m^{2}+\Lambda)}{m^{2}\kappa^2}\sqrt{-g}(\frac{2}{\ell^{2}}-M^{2})\nabla^{0}h_{\mu\nu}^{m},\,\,\,\,\,\,\,\,\,\,\,\,\,\,\,\,\,\,\,\,\,\,\,\,\end{equation}
  \begin{equation}\Pi_{M}^{(1)\mu\nu}=-\frac{2(m^{2}+\Lambda)}{m^{2}\kappa^2}\sqrt{-g}(\frac{2}{\ell^{2}})\nabla^{0}h_{\mu\nu}^{M},\,\,\,\,\,\,\,\,\,\,\,\,\,\,\,\,\,\,\,\,\
  \,\,\,\,\,\,\,\,\,\,\,\,\,\,\,\,\,\,\,\,\end{equation}
    since we have up to four time derivatives in the lagrangian, one can introduce a canonical variable using the Ostrogradsky method \cite{3,4} as $K_{\mu\nu}=\bar{\nabla}_{0}h_{\mu\nu}$. The conjugate momentum of this variable is given by\begin{equation}\Pi^{(2)\mu\nu}=\frac{2(m^{2}+\Lambda)}{m^{2}\kappa^2}\sqrt{-g}\,g^{00}\bar{\Box} h^{\mu\nu},\end{equation}
then using the equations of motion we obtain
  \begin{equation}\Pi_{m}^{(2)\mu\nu}=-\frac{2(m^{2}+\Lambda)}{m^{2}\kappa^2}\sqrt{-g}\,g^{00}(\frac{2}{\ell^{2}})\, h_{m}^{\mu\nu},\,\,\,\,\,\,\,\,\,\,\,\,\,\end{equation}
  \begin{equation}\Pi_{M}^{(2)\mu\nu}=-\frac{2(m^{2}+\Lambda)}{m^{2}\kappa^2}\sqrt{-g}\,g^{00}(\frac{2}{\ell^{2}}-M^{2})\, h_{M}^{\mu\nu},\end{equation}

 Now we can write the Hamiltonian of the system as
  \begin{equation}H=\int d^{2}x(\dot{h}_{\mu\nu}\Pi^{(1)\mu\nu}+\dot{K}_{\mu\nu}\Pi^{(2)\mu\nu}-\pounds),\end{equation}
Then by substituting the equation of motion of the highest weight states, we obtain the energy of massless and massive modes as following
 \begin{equation}E_{m}=\int d^{2}x\,(\dot{h}_{m\,\mu\nu}\Pi^{(1)\mu\nu}_{m}+\dot{K}_{m\,\mu\nu}\Pi^{(2)\mu\nu}_{m}-\pounds)=
 \frac{2(m^{4}+2\Lambda m^{2}+\Lambda^{2})}{m^{2}\kappa^2}\int d^{2}x\sqrt{-g}\dot{h}_{m\,\mu\nu}\bar{\nabla}^{0}h_{m}^{\mu\nu},\end{equation}
 \begin{equation}E_{M}=\int d^{2}x\,(\dot{h}_{M\,\mu\nu}\Pi^{(1)\mu\nu}_{M}+\dot{K}_{M\,\mu\nu}\Pi^{(2)\mu\nu}_{M}-\pounds)=
 -\frac{2(m^{4}+2\Lambda m^{2}+\Lambda^{2})}{m^{2}\kappa^2}\int d^{2}x\sqrt{-g}\dot{h}_{M\,\mu\nu}\bar{\nabla}^{0}h_{M}^{\mu\nu},\end{equation}
  Since the integrals in Eqs.$(5.10)$ and $(5.11)$ are negative, therefore we find that if $m^{2}\ell^{2}<2+\frac{1}{m^{2}\ell^{4}}$, then the energy of the massless modes $h_{\mu\nu}^{m}$ is positive and  it is negative for $m^{2}\ell^{2}>2+\frac{1}{m^{2}\ell^{4}} $. In contrast for the massive modes $h_{\mu\nu}^{M}$ , if $m^{2}\ell^{2}<2+\frac{1}{m^{2}\ell^{4}}$, then the energy  is negative  and it is positive for $m^{2}\ell^{2}>2+\frac{1}{m^{2}\ell^{4}}$.

\section{Conclusion}
In the present paper we have studied the Born-Infeld Gravity in
the $AdS_3$ background. We have obtained the linearized equation
of motion around the $AdS_3$ background. In \cite{5}, the equation
of motion for graviton  have been derived that here we have
obtained the linearized equation of motion and it has been
factorized as a massless second order differential equation and a
massive second order differential equation. We have shown that the Born-Infeld Gravity does not have a critical point for any regular choice of parameters, so in the framework of this model the logarithmic modes do not appear. This behaviuor is in contrast to the NMG, TMG, and GMG models. Therefore Born-Infeld Gravity does not have a logarithmic conformal field theory dual for any regular choice of parameters \cite{20}.
  Then we have obtained the energies of both the highest  massless and massive modes. If $m^{2}\ell^{2}<2+\frac{1}{m^{2}\ell^{4}}$ then the energy of the massless mode is positive and is negative for the massive mode, but if $m^{2}\ell^{2}>2+\frac{1}{m^{2}\ell^{4}}$, the energy of the massless mode is negative and positive for the massive mode. The excitation energies $E_{m}$, $E_{M}$ have opposite sign.
\section*{Acknowledgments}
We thank Aninda. Sinha, Daniel. Grumiller,  Yan. Liu, Bayram. Tekin, and Sang-Heon Yi  for helpful discussions
and correspondence.

\end{document}